\documentclass[12pt,titlepage]{article}
\usepackage{graphicx,amssymb,amsmath}

\begin{document}

\title{Atomic data activities by the OP and IP consortia:
past, present and new perspectives within the VAMDC}

\author{Mendoza$^{1,2}$, Claudio; N\'u\~nez$^{2,3}$, Luis A. \\
$^1$ Centro de F\'{\i}sica, Instituto Venezolano de Investigaciones
Cient\'{\i}ficas (IVIC), \\ PO Box 20632, Caracas 1020A, Venezuela.\\
$^2$ Centro Nacional de C\'alculo Cient\'{\i}fico Universidad de Los Andes
(CeCalCULA), \\ Corporaci\'on Parque Tecnol\'ogico de M\'erida, M\'erida 5101, Venezuela.\\
$^3$ Centro de F\'{\i}sica Fundamental, Facultad de Ciencias,
Universidad de Los Andes, \\ M\'erida 5101, Venezuela.}

\date{February 2010 \\ Review to be published in {\em Acta Cient\'{\i}fica Venezolana} }

\maketitle


\section*{Abstract}

The Opacity Project (OP) and Iron Project (IP) are pioneering
international collaborations which have been computing, for
more than 25 years, massive atomic data sets for astrophysical
applications. We review the data activities that have been
carried out, namely curation, analysis and preservation, and
the development of databases and computer tools for data
dissemination and end-user processing. New opportunities within
the current data-intensive boom referred to as e-science are
described, in particular the Virtual Atomic and Molecular Data
Center (VAMDC) that has been recently launched to consolidate
and promote atomic and molecular database services. {\bf Key
words:} atomic data; Opacity Project; Iron Project; laboratory
astrophysics; databases; e-science; virtual data centers.

\section*{Introduction}

It is an honor to join the celebrations of the 50th Anniversary
of the Instituto Venezolano de Investigaciones Cient\'{i}ficas
(IVIC) by contributing with a review of the data activities of
the Opacity Project (OP) \cite{opt95, opt97} and Iron Project
(IP) \cite{hum93}. Since the beginning of the 80s (OP) and
through the 90s (IP), these international consortia have been
dedicated to the computation of massive atomic data sets for
astrophysical applications. In reference to the title, by data
activities we mean compilations, databases and data curation,
analysis, dissemination and preservation. They have been
supported by the Centre de Donn\'ees astronomiques de
Strasbourg (CDS\footnote{\tt http://cds.u-strasbg.fr/}),
France, the Ohio Supercomputer Center (OSC\footnote{\tt
www.osc.edu/}), USA, the Centro Cient\'{\i}fico de IBM de
Venezuela, the Centro Nacional de C\'alculo Cient\'{\i}fico de
la Universidad de Los Andes (CeCalCULA\footnote{\tt
http://www.cecalc.ula.ve/}) and the Centro de
F\'{\i}sica\footnote{\tt http://www.ivic.ve/fisica/} of IVIC.

It is also timely to pay our respects to Mike
Seaton\footnote{\tt
http://en.wikipedia.org/wiki/M.\_J.\_Seaton}, the leader of the
OP and a central reference to the IP, who passed away in May
2007. As can be appreciated from the long lists of
publications, both the OP\footnote{\tt
http://cdsweb.u-strasbg.fr/topbase/publi.html} and
IP\footnote{\tt
http://cdsweb.u-strasbg.fr/tipbase/ref/publi.html} have been
widely reviewed, most recently in the meeting ``Atoms and
Astrophysics: Mike Seaton's Legacy" held at University College
London, UK, on 14--15 April 2008, to reminisce over his
monumental scientific contributions. A collection of the papers
presented in this event, edited by Pete Storey and Phil Burke
(see Ref.~\cite{sto08} and references therein), acclaims the OP
to be among his major lifetime projects. What perhaps have been
overlooked thus far were his data management skills, views and
endeavors which were crucial in the OP and a source of
encouragement to other atomic data ventures since then. With
him in mind, we will go over the numerical methods and
approximations used in the OP and IP, highlight some of the
main findings, particularly those resulting from data analyzes,
and describe the atomic databases and applications
TOPbase\footnote{\tt
http://cdsweb.u-strasbg.fr/topbase/topbase.html},
TIPbase\footnote{\tt
http://cdsweb.u-strasbg.fr/tipbase/home.html},
OPserver\footnote{\tt http://opacities.osc.edu/} and
CHIANTI\footnote{\tt http://www.chianti.rl.ac.uk/} associated
with these pioneering long-term collaborations.

If data activities were nascent and poorly funded in the 80s
and 90s, they are currently undergoing a tremendous boom as a
new way of doing research, {\em e-science} \cite{hey03,hey05,
fos05}, based on global data-mining collaborations on a second
generation Internet. Due to its foreseeable potential,
data-intensive computing is being put forward as the ``fourth
paradigm" in science \cite{hey09} which intends to unify the
other three: experiment, theory and computer simulation. It
will certainly give rise to new opportunities for
interdisciplinary cooperation and inter-organizational data
sharing where, in the same way that scientific papers are now
globally available on the Internet, so will be the voluminous
distributed data sets that are used to generate them.
Therefore, some of the discussions that are presented here
regarding data management perspectives are of general interest
to other research fields beyond those of computational atomic
physics and laboratory astrophysics.

In this challenging context, the Virtual Atomic and Molecular
Data Center (VAMDC\footnote{\tt http://www.vamdc.org/}) was
launched in July 2009 to integrate and boost atomic and
molecular (A\&M) databases due to their relevance in a variety
of scientific and technological fields. It has been agreed to
house the OP and IP data products and its related
data-intensive applications within the VAMDC and to adopt its
general data-preservation policies, an issue of high priority
since both the OP and IP are now long past their prime time. In
this respect, we will go over recent innovations such as
metadata, XML schema and distributed data reservoirs, and
discuss the perspectives of the OP/IP data services within the
VAMDC and in the new cyber-infrastructure.


\section*{The Opacity Project}

The Opacity Project (OP) \cite{opt95, opt97} was launched as
the result of a plea by Norman Simon \cite{sim82} for a
revision of the metal (elements heavier than He) opacities, and
was first discussed by David Hummer, Dimitri Mihalas and  Mike
Seaton in the living room of the former in Boulder, Colorado,
USA, in the summer of 1982. Simon had shown earlier that an
increase of a factor of 2 to 3 in the metal opacities would
eliminate the long-standing ``Cepheid mass discrepancy".
Cepheids are pulsating variable stars which are used, due to
the precise relationship between their variable luminosity and
pulsation period, to measure extragalactic distances. However,
the Cepheid masses determined from evolutionary tracks and
pulsation properties showed dissatisfying differences
\cite{men90}. The discussion in Boulder dealt with the details
of a new equation of state \cite{hum88}, which would be based
on the ``chemical picture" where atoms keep their identities
and the plasma effects are represented by an occupation
probability formalism, and on the feasibility of including
electron correlation effects in the computation of the massive
radiative data needed to estimate opacities.

Also as a result of Simon's request, a parallel project was
launched at the Lawrence Livermore National Laboratory, CA,
USA, which came to be known as OPAL \cite{igl87}. By using the
method of detailed configuration accounting, where the atomic
data are computed with the Dirac equation and parametric
potentials, and an equation of state \cite{rog86} based on the
``physical picture" by a renormalized activity expansion of the
grand canonical ensemble, OPAL removed some of the inherent
approximations in the widely used Los Alamos Astrophysical
Opacity Library \cite{hue77}.

The OP consortium quickly grew to involve about 30
international collaborators\footnote{\tt
http://cdsweb.u-strasbg.fr/topbase/OPteam.html} from France,
Germany, UK, USA and Venezuela, and would meet regularly on a
six-monthly basis for at least the next ten years. In a way,
the OP established a milestone in the field of computational
laboratory astrophysics regarding participation in long-term
international scientific collaborations, publication of results
(e.g. the series of papers\footnote{\tt
http://cdsweb.u-strasbg.fr/topbase/publi.html} ``The equation
of state for stellar envelopes" and ``Atomic data for opacity
calculations") and data management, the latter being the main
topic of the present review.

Radiation transfer in the stellar envelope, with typical
temperatures $T$(K) and densities $\rho$(g~cm$^{-3}$) in the
ranges
\begin{equation}
4.5\leq \log T\leq 6.5 \quad\quad \hbox{\rm and} \quad  -8.5\leq\log\rho\leq -1.5\ ,
\end{equation}
can be modeled in local thermodynamic equilibrium. Hence, the
specific intensity is not very different from the Planck
function $B(T)$, and the total radiant energy flux is assumed
proportional to the temperature gradient (diffusion
approximation):
\begin{equation}
{\bold F} = -\left(\frac{4\pi}{3\kappa_{\rm R}}\frac{{\rm d}B}{{\rm d}T}\right)
\nabla T \ .
\end{equation}
The transfer is thus controlled by the {\em Rosseland mean}
\begin{equation}
\label{rmo}
\frac{1}{\kappa_{\rm R}} = \int_0^\infty \frac{1}{\kappa_\nu}\times g(u){\rm d}u
\end{equation}
where $g(h\nu/kT)$ is the weighting function
\begin{equation}
\label{wfun}
g(u) = \frac{15}{4\pi^4}u^4\exp(-u)[1-\exp(-u)]^{-2}\ ,
\end{equation}
and the {\em monochromatic opacities} $\kappa_\nu$ include
contributions from all the radiative absorption processes in
the plasma. It must be pointed out that the Rosseland mean is a
weighted harmonic mean that is not additive, i.e. all the
contributions to the monochromatic opacity must be added up
before the integral in Eqn.~(\ref{rmo}) is computed. It also
weighs the windows between absorption lines rather that the
strong absorption lines themselves, and therefore, accurate
representations must be considered for the line wings, the weak
lines and absorption lines originating from minor constituents
of the plasma.

The required radiative atomic data, namely level energies,
$f$-values, photoionization cross sections and line broadening
parameters, were computed in the OP with the {\em
close-coupling method} \cite{bur71b}, where the wave function
for a state of an ionic $N$-electron target and a colliding
electron with total orbital angular momentum, spin and parity
$SL\pi$ is expanded in terms of the target eigenfunctions
\begin{equation}\label{cc}
\Psi(SL\pi) ={\cal A}\sum_i \chi_i{F_i(r)\over r}+\sum_jc_j\Phi_j\ .
\end{equation}
The functions $\chi_i$ are vector coupled products of the
target eigenfunctions and the angular components of the
incident-electron functions, $F_i(r)$ are the radial part of
the latter and $\mathcal{A}$ is an antisymmetrization operator.
The functions $\Phi_j$ are bound-type functions of the total
system constructed with target orbitals, introduced to
compensate for orthogonality conditions imposed on the $F_i(r)$
and to improve short-range correlations. The Kohn variational
principle gives rise to a set of coupled integro-differential
equations which were solved in the OP with the {\em R-matrix
method} \cite{ber87, ber74, ber78, bur71a} and a series of
sophisticated asymptotic codes developed by Seaton \cite{sea86}
to improve and accelerate the calculation of bound-state and
continuum energies and wave functions, and hence, the radiative
data. For instance, it became possible, for the first time, to
obtain all possible bound states in an energy range by scanning
rather than iterating around initial energy guesses. It is
worth emphasizing, on the other hand, that in the OP $R$-matrix
computations relativistic effects were neglected and $LS$
coupling was assumed, so fine-structure splittings were not
resolved. The rationale behind this approximation was that line
broadening would blur such splittings at the expense of
neglecting the weak intercombination transitions that arise
from relativistic spin mixing.

Data computations proceeded by assigning complete isoelectronic
sequences to specific research groups which had to report
periodically on their progress at OP meetings. Atomic data
curation in the OP was mostly managed on half-inch tapes and,
in some cases, on the newer Exabyte 8-mm cassettes and CDs and
was centralized by Seaton. He developed stringent testing
procedures in order to ensure data accuracy and completeness
and to remove nonphysical resonances in the photoionization
cross sections, particularly in the difficult energy regions
below thresholds. In many cases calculations were repeated due
to programming bugs or poor target representations. The final
opacity computations \cite{sea94} were performed in the
difficult transition period between the twilight of the
mainframes and the advent of the Unix workstations, so they
implied familiarity with different operating systems (IBM
VM/CMS, VAX/VMS and Unix) and fortran compilers, the end
product being a CD that was publicly available from two sites:
the Observatoire de Meudon, Paris, France, and the Department
of Astronomy, Ohio State University, Columbus, OH, USA.

One of the most interesting findings in the OP was obtained by
data analysis, namely in the photoionization cross sections of
excited states. The most salient features in such cross
sections were broad resonances that involved the
photoexcitation of the core rather than that of the active
electron, and were consequently labeled {\em PEC resonances}.
In Figure~\ref{fernley_etal_etal}, we show the OP
photoionization cross sections \cite{fer87} of the ${\rm
1s}n{\rm s}\ ^1{\rm S}$ excited states in He~I for $n=2{-}4$,
where radiative absorption can give rise to the excitation of a
series of ${\rm 2p}n'{\rm s}$ resonances converging to the
${\rm 2p}$ threshold
\begin{equation}
{\rm 1s}n{\rm s} + \gamma \rightarrow {\rm 2p}n'{\rm s}
\rightarrow {\rm 1s} + {\rm e^-}\ .
\end{equation}
The PEC resonances arise when $n'=n$ since the active electron
does not participate in the radiative excitation which takes
place solely via the ${\rm 1s}\rightarrow {\rm 2p}$ core
transition; therefore, the PEC resonance widths are mostly
determined by this core process and are then practically
independent of $n$.


\begin{figure}
\centering
\caption{\label{fernley_etal_etal} OP photoionization cross sections
of the ${\rm 1s}n{\rm s}\ ^1{\rm S}$ excited states ($n=2{-}4$)
in He~I as a function of the effective quantum number $E=-1/\nu^2$.
It may be seen that the cross section of each state is dominated
by a broad PEC resonance which is assigned the term
${\rm 2p}n'{\rm s}\ ^1{\rm P^o}$ with $n'=n$. In PEC resonances,
the active electron does not take part in the radiative absorption,
and hence, its width is determined by the ${\rm 1s}\rightarrow {\rm 2p}$
core transition. Reproduced from Figures~7 of Ref.~\cite{fer87}
(http://dx.doi.org/10.1088/0022-3700/20/23/032).}
\end{figure}


In the lengthy revisions, the first preliminary opacities were
not published \cite{igl87} by OP but by OPAL, a comparison of
the Los Alamos and OPAL monochromatic opacities for Fe which we
show in Figure~\ref{iglesias_etal}. A huge bump may be seen at
around 60~eV in OPAL which is absent in Los Alamos; it is due
to an unresolved transition array involving numerous states of
the type 3s$^x$3p$^y$3d$^z$ in ionic species of the second row,
namely in Al-, Si-, P-, S- and Cl-like Fe. As mentioned by the
authors, this bump would greatly enhance the Rosseland mean
since its distribution function (see Eqn.~\ref{wfun}) peaks at
around 80~eV. Furthermore, this outcome became a key milestone
in the opacity race; firstly, it was an incentive for the two
competing teams since Simon's hypothesis appeared to be correct
and new revised opacities were indeed required; and secondly,
the large number of 3s$^x$3p$^y$3d$^z$ states could not be
handled computationally at the time with the $R$-matrix
approach. Within the OP, the latter situation led to additional
extensive calculations of $f$-values with the atomic structure
code SUPERSTRUCTURE \cite{eis74} which uses simpler
configuration-interaction wave functions of the type
\begin{equation}
\Psi(SL\pi)=\sum_i c_i \phi_i\ .
\end{equation}
These new large data sets came to be known as the ``PLUS data"
\cite{lyn95}.

In spite of using very different approaches for the equation of
state and the computation of the atomic data, the independent
OP and OPAL opacities turned out to be, after more than a
decade of heavy computations, in outstanding agreement and have
been extensively compared. The first comparisons were performed
of course in the context of pulsation calculations, namely in B
stars \cite{pam94} and beat and bump Cepheids and RR Lyrae
stars \cite{kan94}, where the new OPAL and OP opacities
generally contributed to improved and undifferentiated models.
At temperatures of $\sim$10$^5$~K characteristic of stellar
interiors, the OPAL and OP opacities were in close agreement
\cite{bad03} and larger than previous values by as much as a
factor of 3. However, at higher temperatures and densities,
such as those found in the deeper layers, OPAL was higher than
OP by around 30\% which was suggested \cite{igl95} to be due to
the neglect of inner-shell transitions in the latter. This
proposition was subsequently confirmed \cite{bad03} by
computing new inner-shell radiative data for a mixture of six
elements (H, He, C, O, S and Fe) with AUTOSTRUCTURE
\cite{bad86, bad97}, an extended version of the atomic
structure code SUPERSTRUCTURE \cite{eis74}.


\begin{figure}
\centering
\caption{\label{iglesias_etal} Comparison of Los Alamos monochromatic opacities
for Fe with OPAL. Left panel: photon absorption coefficient by Los Alamos code
(1985 version) at a density of $6.82\times 10^{-5}$ g~cm$^{-3}$ and a temperature
of 20~eV. Right panel: OPAL results where a huge unresolved transition array is
observed at around 60~eV due to numerous 3-to-3 transitions not taken
into account by Los Alamos. Reproduced from Figures~1--2 of Ref.
\cite{igl87} (http://dx.doi.org/10.1086/185034).}
\end{figure}


A more detailed comparison \cite{sea04} of OPAL and OP opacites
yielded some significant differences. For instance, although
there was good global agreement for hydrogen, in the region of
$\log T\simeq 6$ ($T$ in K) and $\log R\simeq -1$, where $R
=\rho/T^3_6$ ($\rho$ is the mass density in g~cm$^{-3}$ and
$T_6 = 10^{-6}\times T$), the OPAL Rosseland mean was larger
than OP by up to 13\% which was mainly caused by the different
equations of state. (For a discussion on the validity of the
two equations of state, see Ref.~\cite{dap87, tram06}.) Similar
differences occurred for He at $\log T\simeq 6.4$ and $\log R
\simeq -1$. It was also found that the inclusion of
intercombination transitions in iron increased the Rosseland
mean by 18\%, and in iron-rich mixtures, the OP ``$Z$-bump" was
located at somewhat higher temperatures when compared to OPAL.
Generally speaking, there are small differences between OPAL
and OP at the lower temperatures ($\log T\lesssim 5.5$) which
are believed to be due to atomic data quality, OP certainly
having the upper hand. At the higher temperatures and densities
($\log R\gtrsim -2$), OP is somewhat larger due to a different
equation of state.

These conclusions led to a major revision of the OP opacities
\cite{bad05}. The outer-shell iron data \cite{lyn95, sea04}
obtained in intermediate-coupling with AUTOSTRUCTURE for ionic
species with electron number $N = 13$--18 were now included,
plus inner-shell radiative data for the chemical elements He,
C, N, O, Ne, Na, Mg, Al, Si, S, Ar, Ca, Cr, Mn, Fe and Ni. An
improved frequency mesh was also introduced to ensure a high
degree of accuracy in integration (better than 1\%). The
contributions of inner-shell transitions may be appreciated in
Figure~\ref{badnell_etal1} in the high-temperature ($\log T
\gtrsim 5.5$), high-density ($\log R\gtrsim -3$) tails.
Moreover, as shown in Figure~\ref{badnell_etal2}, the OP
revised opacities agreed with OPAL to better than 10\%, but
again, the OP $Z$-bump at $\log T \simeq 5.2$ was shifted to
slightly higher temperatures.

An interesting development took place while the OP opacities
were being improved. By means of a time-dependent,
three-dimensional, hydrodynamical model \cite{asp04} of the
solar atmosphere, the C, N, O and Ne solar abundances were
revised downward to yield a photospheric metal mass fraction of
$Z=0.0126$, a value considerably lower than the widely accepted
value of $Z=0.0194$ derived \cite{and89, gre98} from standard
solar abundances. The new estimates not only sadly destroyed
the outstanding agreement between solar models and
helioseismological observations in relation to the depth of the
solar convection zone, respectively 0.714~R$_\odot$ and
$0.713\pm 0.001$~R$_\odot$, with a larger value
(0.726~R$_\odot$), but it also led to discrepancies with the
measured sound speeds and the surface helium abundance
\cite{bah05}. It was suggested \cite{bah05} that an increase of
11\% in the OPAL opacities from the base of the convection zone
($0.7$~R$_\odot$) down to $0.4$~R$_\odot$ would solve the
problem. Thus, there was much expectation for the OP revision
to confirm this hypothesis, but the final OP increase was less
than 2.5\% more than OPAL in the region of interest
\cite{bad05}. Possible sources of ``missing opacity" in the
context of the solar abundances problem have been recently
proposed \cite{pra09}, where the neglected intercombination
transitions and relatively simple target representations used
(only the ground complex was considered in most cases) in the
OP are top of the list.


\begin{figure}
\centering
\caption{\label{badnell_etal1} OP Rosseland-mean opacities for a solar mixture (S92)
with and without inner-shell transitions. Reproduced from Figure~1 of Ref.
\cite{bad05} (http://dx.doi.org/10.1111/j.1365-2966.2005.08991.x).}
\end{figure}


\begin{figure}
\centering
\caption{\label{badnell_etal2} OPAL and OP Rosseland-mean opacities for a solar
mixture (S92) where an excellent overall agreement is found. Reproduced from
Figure~2 of Ref. \cite{bad05} (http://dx.doi.org/10.1111/j.1365-2966.2005.08991.x).}
\end{figure}


The computation of vast numbers of $f$-values, photoionization
cross sections, line broadening parameters and monochromatic
opacities in the OP allowed the determination of radiative
accelerations which can be used to study diffusion processes in
stars \cite{ale95}. Diffusion can be the cause of surface
abundance anomalies in, for instance, chemically peculiar stars
(e.g. HgMn stars). Seaton \cite{sea96} used the OP radiative
data to study the phenomenon of {\em radiative levitation}  and
to compute envelope radiative accelerations for 15 elements
\cite{sea97}, making publicly available a useful suite of
fortran utilities for estimating radiative acceleration
parameters. He also treated in detail \cite{sea99} the envelope
diffusion of iron-group elements in HgMn stars which produced
changes in the Rosseland-mean opacities by as much as a factor
of 4, particularly producing build-ups of iron in the region of
$\log T \simeq 5.1$ and of nickel in the outer parts of the
envelope. A comparison of Mn abundances obtained from
observations in the UV \cite{smi93} and optical \cite{ade96} of
HgMn stars with those calculated at the bottom of a model
atmosphere (optical depth $\tau =1$) and after $10^8$ years
showed satisfactory agreement. It has been argued \cite{del08}
that radiative accelerations would enable further tests of OP
and OPAL opacities by comparing such observed surface abundance
anomalies with those predicted in diffusive stellar models.
Significant differences in the OP and OPAL accelerations have
been recently discussed \cite{pra09}.

\section*{TOPbase}

Most of the work by one of us (CM) in the OP was carried out
while he was a Research Consultant at the IBM Venezuela
Scientific Center and a Visiting Fellow of IVIC. This allowed
him to be in contact with other scientists involved in
scientific computing including computer scientists. Thus, when
the bulky OP atomic data sets were finally being gathered and
tested, he was encouraged by this circle to become acquainted
with the field of {\em data dissemination}.

In those days, laboratory astrophysicists were familiar with
the atomic data compilations published by the US National
Bureau of Standards (now the National Institute of Standards
and Technology), of which the most well-known were {\em Atomic
Energy Levels} \cite{moo49} and {\em Atomic Transition
Probabilities} \cite{wie66}, but there was scanty contact with
online atomic databases or commercial database management
systems (DBMSs). Early experiences, e.g. the Belfast Atomic
Data Bank \cite{ber91}, were mainly concerned with data
reservoirs and bibliographies for data assessment within the
context of atomic data workshops \cite{eis86, eis88, zei91}
rather than with user data products.

Thus, taking into consideration that the OP atomic data were
likely to be employed in a variety of research fields and for
different purposes, there was at the time a definite motivation
to enhance their access and usage by developing a versatile and
searchable atomic database. However, the diversity of
scientific environments, resources and operating systems
implied, right from the ouset, that the proposed DBMS would
have to be custom-developed in the scientific computing {\em
lingua franca}: fortran 77. In this respect, we were able to
count with the enthusiastic support of Walter Cunto, a first
class computer scientist at the IBM Scientific Center, who was
responsible for most of the design of what became to be known
as ``TOPbase" \cite{cun92}, i.e. the definition, development,
distribution and maintenance of a specific, portable, efficient
and low-cost DBMS to facilitate the intensive use (online and
batch) of the OP atomic data sets.

The original version of TOPbase \cite{cun92} handled around
0.5~GB of compact binary data files, organized to speed up
searches along spectroscopic series and isonuclear and
isoelectronic sequences and the sorting of energies and
wavelengths. User searches were specified by means of a simple
command-based query language which included graphic processing.
Its structure essentially contained three entities --- term
energies ($e$), $f$-values ($f$) and photoionization cross
sections ($p$) --- and a set of indexes, all resident in
secondary storage (disk). When TOPbase was invoked, the indexes
were loaded into main memory (RAM) such that the query parser
could use them to streamline data searches. Index displays
could also be employed to obtain tables of content of the
database. The main bottleneck in system performance was, due to
the data volumes involved, data loading from secondary storage.
As a consequence, the data manipulation scheme was implemented
at two levels: (i) searches in and time consuming block data
retrievals from secondary storage; and (ii) cheap and versatile
processing (sorting, row/column selection, exclusion and
plotting) performed iteratively in main storage in order to
satisfy the user's ultimate requirements.


\begin{figure}
\centering
\includegraphics[width=12cm]{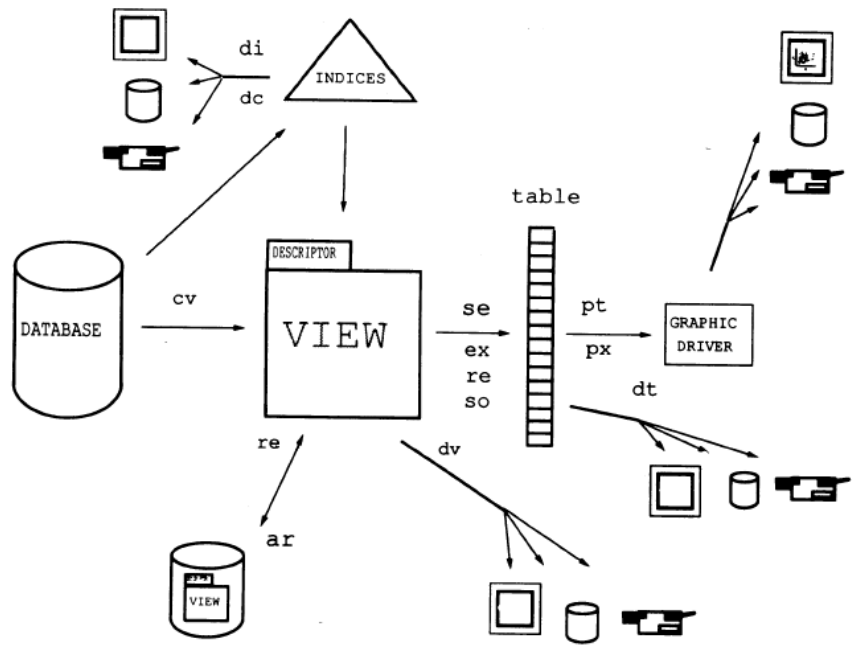}
\caption{\label{cunto_mendoza} TOPbase data model showing the two
main data structures, the {\em view} and  {\em table}, the display
and graphic capabilities  and query commands.
Reproduced from Figure~2 of Ref. \cite{cun92} with permission of
$^\copyright$RevMexAA.}
\end{figure}


The TOPbase data model is shown in Figure~\ref{cunto_mendoza}.
Data compactness and fast access were two main features whereby
main and secondary storages were managed jointly with the
logical handling of the data. Two data structures were
therefore implemented in main storage: the {\tt view} and the
{\tt table}. A search was performed according to a user
selected criterion that generated a subset of highly cohesive
data, i.e. a {\tt view}, which was loaded into special buffers
located in main memory with the {\tt cv} command to allow
further manipulation. Each {\tt view} had an associated {\tt
descriptor} to register selection criteria and {\tt view}
bounds which could be displayed on the monitor at any time. A
{\tt view} binary image could be respectively archived in or
retrieved from disk with the {\tt ar} and {\tt re} commands,
and the package offered data streaming facilities to different
output devices, namely the monitor, printer and a disk file
through the {\tt dv} command. Logical reorganizations of data
stored in a {\tt view} were possible through the {\tt table}
structure. A {\tt table} was a vector array that enabled or
disabled data items contained in the {\tt view} as a result of
user selection requirements such as {\tt se} (selection), {\tt
ex} (exclusion) and {\tt so} (sorting). A {\tt table} could
also be output with the {\tt dt} command on different devices
(monitor, printer or disk file), and graphic displays of its
columns ({\tt pt}) and photoionization cross sections ({\tt
px}) could be plotted. Tables of content of the database, which
were useful before creating {\tt views}, could be obtained by
displaying  the indexes ({\tt di}), and a set of atomic
constants (e.g. $Z$-dependent Rydberg constants) were also
available with the {\tt dc} command.

Although TOPbase was originally developed such that it could be
installed locally on any system with a fortran~77 compiler and
distributed on tape, we soon became aware that in academic
environments the rapidly spreading Internet allowed remote
access in addition to the popular electronic mail. Thus a more
efficient scheme was to install the database at a central site,
from which it could be invoked online with network commands
such as {\tt telnet} and {\tt ftp}. This arrangement was
enthusiastically promoted by Claude Zeippen (an active OP
participant at the Observatoire de Paris, France) who proposed
the Centre de Donn\'ees astronomiques de Strasbourg
(CDS\footnote{\tt http://cds.u-strasbg.fr/}) as a convenient
host. With much appreciated technical assistance from
Fran\c{c}ois Ochsenbein, TOPbase became operable \cite{cun93}
at the CDS in January 1993 and has remained so up to the
present day.

With the appearance of the World Wide Web in the mid 90s,
TOPbase became a handy case study in the new technology. A
web-based user interface was developed in 1995 by one of us
(CM) at IVIC with the assistance of Jes\'us Quiroz, and
included soon after in the CDS home page. The current TOPbase
web service\footnote{\tt
http://cdsweb.u-strasbg.fr/topbase/topbase.html} is an abridged
version of the command-based system described above,
particularly the versatile {\tt table} structure was not
implemented due to difficulties in postprocessing data after
they have been streamed by the web browsers. The Web allows
estimates of access frequency, and in the case of TOPbase, the
General Internet Search Engine for Atomic Data
(GENIE\footnote{\tt http://www-amdis.iaea.org/GENIE/}) provided
consistent monthly statistics of over 100 hits in the last few
years.


\section*{OPserver}

In the early days of TOPbase development at the IBM Venezuela
Scientific Center, lengthy discussions with Walter Cunto on
atomic databases \cite{cun92} always considered two access
modes: interactive online access by a user and recurrent access
in batch processing by an application. Although the former mode
has grown since then, particularly with the ubiquitous
establishment of the Web, the latter has remained in the
background. However, in the new cyber-infrastructure of
e-science, characterized by the mining of large volumes of data
in distributed reservoirs, it is expected to take a leading
role. In this same context, we predict that the deployment of
data-intensive applications, which at present implies a local
download, installation and periodic upgrading, will change at
its roots.

The development of the OPserver \cite{men07} was carried out
with these ideas in mind, but above all, with the contributions
of several students and postdocs: Marcio Mel\'endez
(Universidad Sim\'on Bol\'{\i}var, Caracas), Juan Gonz\'alez
and Enrique Palacios (Universidad de Carabobo, Valencia), Luis
Rodr\'{\i}guez (IVIC, Caracas), Franck Delahaye (Observatoire
de Paris), and the technical support, expertise and
encouragement of Paul Buerger (Ohio Supercomputer Center,
Columbus), Alberto Bellor\'{\i}n (Universidad Central de
Venezuela, Caracas), Anil Pradhan (Ohio State University,
Columbus), Claude Zeippen (Observatoire de Paris) and Mike
Seaton himself.

With the last revision \cite{bad05} of the OP opacities to
treat the inner-shell contribution, a complete set of
monochromatic opacities and a suite of simple-to-use codes to
compute means and radiative accelerations
(OPCD\_2.1\footnote{\tt
http://cdsweb.u-strasbg.fr/topbase/op.html}) was released
\cite{sea05}. This was, in our opinion, a major achievement as
it represented the culmination of a 20-year endeavor embodied
in a user data product. However, as mentioned before (see the
section on The Opacity Project), some sophisticated stellar
models take into account microscopic diffusion processes
(radiative levitation, gravitational settling and thermal
diffusion) which means that chemical element stratification
depends on stellar depth. This implies that mean opacities and
radiative accelerations must be recalculated at each depth
point of the model and at each time step of the evolution, and
thus codes more efficient than OPCD\_2.1 may be necessary.
OPserver attempted to enhance OPCD\_2.1 by: (i) storing the
monochromatic opacities permanently in main memory; (ii)
transcribing the codes as a subroutine library that can be
linked to a stellar modeling code; and (iii) offering several
access modes.

In OPCD\_2.1, Rosseland-mean opacities and radiative
accelerations are computed in two stages (see
Figure~\ref{mendoza_etal1}). In a time-consuming stage and for
a user-defined chemical mixture, the MIXV and ACC codes read
the monochromatic opacities from disk ($\sim$1~GB) and compute
means and accelerations on a representative tabulation of the
complete $T-N_{\rm e}$ plane, where $T$ is the temperature and
$N_{\rm e}$ the electron density. In a second fast stage, the
means and accelerations are interpolated with OPFIT and ACCFIT
on a stellar depth profile specified by the user. In OPserver,
these computations are improved by having a dedicated server in
the first step, where the monochromatic opacities are always
stored in RAM, and by parallelizing the loop over the chemical
mixture.

In Figure~\ref{mendoza_etal2}, we show the access modes offered
by the OPserver which has been resident at the Ohio
Supercomputer Center (OSC), Columbus, OH, USA, since 2006. It
has been implemented under a client--server architecture where
the web server (client) communicates with a supercomputer
(server) via a socket interface. The 1~GB monochromatic opacity
data sets are always loaded in main memory on the server.
OPserver may be accessed by a user (client) from the
interactive web server\footnote{\tt http://opacities.osc.edu}
(Mode~C); by downloading the package locally from the CDS
(OPCD\_3.3\footnote{\tt
http://cdsweb.u-strasbg.fr/topbase/op.html}), installing it and
then linking the subroutine library (OPlibrary) to the user
modeling code (Mode~A); or by just downloading and linking the
OPlibrary, and then accessing the data remotely at runtime from
the central facility at the OSC (Mode~B). The latter mode has
been designed for the new distributed grid computing
environments\footnote{\tt http://www.gridcomputing.com/} where
Mode~A would be cumbersome due to data volumes and deployment
times.

The OPserver has recently been  tested in innovative laboratory
opacity experiments \cite{bai09}, where a sample is heated with
X rays and the spectral transmission measured with a back-light
in a high energy density facility. These initiatives have been
motivated by the solar abundance problem described above (see
the Opacity Project section) and due to the dependence on
opacities from light chemical elements in inertial fusion and
$Z$ pinches.


\begin{figure}
\centering
\caption{\label{mendoza_etal1} Rosseland-mean opacities (RMO) and
radiative accelerations (RA) are computed with the codes in the
OPCD\_2.1 release in two stages: in a time consuming Stage~1, they
are computed for the whole $(T,N_e)$ plane, followed by fast bicubic
interpolations in Stage~2. Reproduced from Figure~1 of Ref.
\cite{men07} (http://dx.doi.org/10.1111/j.1365-2966.2007.11837.x).}
\end{figure}


\begin{figure}
\centering
\caption{\label{mendoza_etal2} OPserver enterprise showing the
web-server--supercomputer tandem at the Ohio Supercomputer Center
(OSC) and the three available user modes. Reproduced from
Figure~2 of Ref. \cite{men07} (http://dx.doi.org/10.1111/j.1365-2966.2007.11837.x).}
\end{figure}


\section*{The Iron Project}

By the end of the OP, most of the researchers that had been
involved in this consortium felt that they wanted to go on, to
get involved in a second project of comparable scale and
relevance. The proposed challenge, the Iron Project (IP)
\cite{hum93}, was just as ambitious: the systematic computation
of accurate radiative and electron impact excitation rates for
ions of the iron group of the Periodic Table, which were
required to interpret a wide variety of astronomical spectra
where local thermodynamic equilibrium in the emitting plasmas
cannot be assumed. The IP work has been published in a long
series of over 60 papers (``Atomic data from the IRON
project"\footnote{\tt
http://cdsweb.u-strasbg.fr/tipbase/ref/publi.html}) involving
the old OP research groups plus new collaborating
institutions\footnote{\tt
http://www.usm.lmu.de/people/ip/institutions.html} (specially
from Canada) under the general coordination of David Hummer.

A main difference with OP was that fine-structure levels would
now have to be  considered, i.e. atomic data in intermediate
coupling would become the end products. In the case of
radiative rates for both allowed and forbidden transitions, the
familiar structure codes SUPERSTRUCTURE \cite{eis74} and CIV3
\cite{hib75} could compute accurate fine-structure $A$-values
for light elements ($Z < 36$, say) with
configuration-interaction expansions of the type
\begin{equation}
\Psi (J\pi) = \sum_i c_i \phi_i
\end{equation}
and by introducing relativistic corrections with the
Breit--Pauli Hamiltonian
\begin{equation}
  \label{hbp}
  H_{\rm bp} = H_{\rm nr} + H_{\rm 1b} + H_{\rm 2b}
\end{equation}
where $H_{\rm nr}$ is the usual non-relativistic Hamiltonian.
The one-body relativistic operators
\begin{equation}
   \label{h1b}
   H_{\rm 1b} = \sum_{n=1}^{N} {f_n({\rm mass}) + f_n({\rm d})
          + f_n({\rm so})}
\end{equation}
represent the spin--orbit interaction, $f_n({\rm so})$, the
non-fine-structure mass variation, $f_n({\rm mass})$, and the
one-body Darwin correction, $f_n({\rm d})$. The two-body Breit
operators are given by
\begin{equation}
   \label{h2b}
   H_{\rm 2b} = \sum_{n<m} g_{nm}({\rm so}) + g_{nm}({\rm ss}) +
   g_{nm}({\rm css}) + g_{nm}({\rm d}) + g_{nm}({\rm oo})
\end{equation}
where the fine-structure terms are $g_{nm}(so)$
(spin--other-orbit and mutual spin--orbit), $g_{nm}(ss)$
(spin--spin), and the non-fine-structure counterparts
$g_{nm}(css)$ (spin--spin contact), $g_{nm}(d)$ (two-body
Darwin) and $g_{nm}(oo)$ (orbit--orbit).

On the other hand, the electron impact excitation rate for a
transition $i\rightarrow f$ is conveniently given in terms of
the dimensionless {\em effective collision strength}
\begin{equation}
\Upsilon_{if}(T) = \int_0^\infty\Omega_{if}(E_f)\exp(-E_f/kT) d(E_f/kT)
\end{equation}
which constitutes the thermal average of the {\em collision
strength} for the transition, namely
\begin{equation}\label{omega}
\Omega_{if} = \frac{1}{2} \sum g |S_{if} -\delta_{if}|^2
\end{equation}
where ${\bf S}$ is the {\em scattering matrix} containing the
micro-physics of the scattering and $g=2J+1$ is a statistical
weight. The excitation cross section may be written in terms of
the collision strength as
\begin{equation}
\sigma_{if} = \Omega_{if}\frac{\pi a_0^2}{g_ik_i^2} \ ,
\end{equation}
$k_i^2$ being the incident electron energy in Rydberg units and
$g_i=2J_i+1$, and the complex scattering matrix is usually
derived in terms of the real {\em reactance matrix}
\begin{equation}
\boldmath{S} = \frac{\boldmath{1} +
         {\rm i}\boldmath{R}}{\boldmath{1} - {\rm i}\boldmath{R}}
\end{equation}
which is in fact the quantity that is actually computed. At the
time of the IP launch, the way to obtain ${\bf R}$ matrices in
intermediate coupling was to compute them in $LS$ coupling, and
then perform the algebraic transformation
\begin{equation}
\label{jajom}
\begin{split}
R^{J\pi} (S_iL_iJ_il_iK_i, S_fL_fJ_fl_fK_f) = &
 \sum_{SL} R^{SL\pi}(S_iL_il_is_i,S_fL_fl_fs_f) \\
 & C(SLJ,S_iL_iJ_i,l_iK_i)C(SLJ,S_fL_fJ_f,l_fK_f)
 \end{split}
\end{equation}
with a code such as JAJOM \cite{sar72, sar78}.
$C(SLJ,S_iL_iJ_i,l_iK_i)$ are Clebsch--Gordon coefficients and
$S_iL_iJ_i$ and $SLJ$ are the orbital angular momentum, spin
and total angular momentum quantum numbers of the target (in
state $i$) and of the total system (target + electron),
respectively; $l_i$ and $s_i$ are the orbital angular momentum
and spin quantum numbers of the electron, and ${\bf K}$ is an
intermediate quantum number such that
\begin{equation}
\boldmath{K_i}  = \boldmath{J_i} + \boldmath{l_i}\quad\quad
   \hbox{\rm and} \quad\quad
   \boldmath{J}  = \boldmath{K_i} + \boldmath{s_i} \ .
\end{equation}
Some allowance for relativistic effects was made \cite{hum93}
by the formalism of {\em term-coupling coefficients} and by
including the non-fine-structure mass variation and one-body
Darwin corrections of Eq.~(\ref{h1b}) in the Hamiltonian of the
$R$-matrix method. Some of the earlier work in the IP was
carried out with this approach; however, a more formal
treatment of the relativistic close-coupling approximation was
needed for some of the complex Fe ionic species.

Although the Breit--Pauli relativistic Hamiltonian had long
been included in $R$-matrix code \cite{sco80,sco82}, the
two-body Breit corrections of Eq.~(\ref{h2b}) were not taken
into account. Nevertheless, by merging this improved
relativistic version with the OP non-relativistic package, an
efficient Breit--Pauli $R$-matrix suite (BPRM) began to be
generally used in the IP as from 1995 for both collisional and
radiative work; e.g. in the electron impact excitation of the
${\rm 3d}^4\ ^5{\rm D}_J$ ground-state fine-structure
transitions in Ti-like ions \cite{ber95}. It was found that,
for the ${\rm 3d}^4\ ^5{\rm D}_0 \rightarrow {\rm 3d}^4\ ^5{\rm
D}_1$ transition in Fe~V, the BPRM effective collision strength
at $10^4$~K was 35\% below that computed with JAJOM. However,
it was soon realized that BPRM led to computationally much more
demanding runs where the number of levels of a target
representation could be limited by processor core size, thus
making some of the complex Fe species computationally
intractable. For this reason, the {\em Intermediate Coupling
Frame Transformation method} (ICFT) was developed \cite{gri98}
(outside the IP but quickly incorporated), whereby relativistic
effects may be included via frame transformations based on the
multi-channel quantum defect theory. For instance, for O-like
Fe~XIX, the original IP calculation \cite{but01} with BPRM
included 92 levels in the target representation while a revisit
\cite{but08} with ICFT allowed up to 342 target levels.

Progress in the IP was translated with time into many
improvements to the $R$-matrix codes and the establishment of
useful numerical utilities in computational atomic physics. For
instance, since the infinite summation of Eq.~(\ref{omega})
converges very slowly for allowed transitions, a ``top-up"
procedure based on the Burgess sum rule \cite{bur86} was
incorporated in BPRM. The inclusion of the two-body Breit
corrections (see Eq.~\ref{h2b}) has also been recently carried
out by G. X. Chen, W. Eissner \& A. K. Pradhan (in
preparation). Furthermore, a graphical method \cite{bur92} for
analyzing the partial-wave convergence of effective collision
strengths based on the reduction of the infinite temperature
scale to a finite interval ($0\leq T\leq 1$) became popular
among IP members. However, it relied on the
infinite-temperature value of the effective collision strength
which, for allowed transitions, may be easily estimated from
the $f$-value. For forbidden transitions, on the other hand, it
is estimated from the Coulomb--Born limit \cite{bur97} which
was not available in the earlier years of the project, and was
eventually coded into SUPERSTRUCTURE by W. Eissner
(unpublished).


\begin{figure}
\centering
\includegraphics[width=10cm]{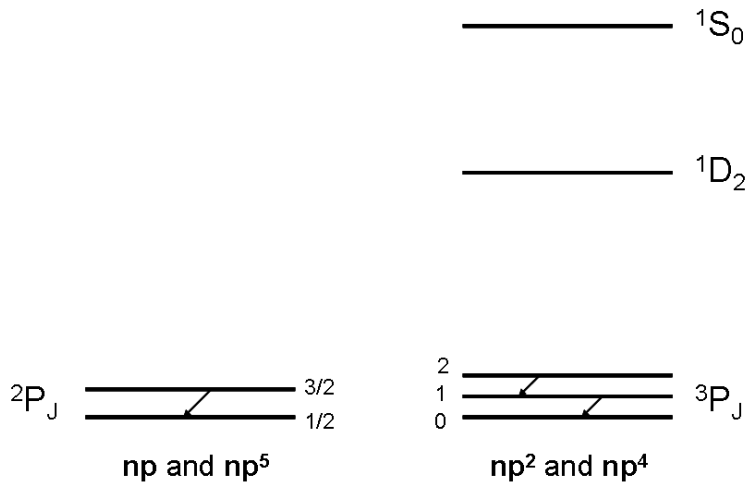}
\caption{\label{IR_lines} Infrared lines originating from ions
with open-shell configurations $n{\rm p}$,  $n{\rm p}^5$,
$n{\rm p}^2$ and $n{\rm p}^4$ respectively belonging, for
$n=2$, to  the B, F, C and O isoelectronic sequences, and for
$n=3$, to the Al, Cl, Si and S sequences.}
\end{figure}


\begin{figure}
\centering
\caption{\label{butler_zeippen} Collision strength for the
${\rm 2s}^2{\rm 2p}^4\, ^3{\rm P_0}\rightarrow\, ^3{\rm P_1}$ infrared
transition in O-like Al~VI displaying a bundle of broad resonances just
above threshold. Reproduced from Figure~2 of Ref.~\cite{but94}
(http://adsabs.harvard.edu/abs/1994A\%26AS..108....1B).}
\end{figure}


The first stage of the IP was concerned with the calculation of
electron impact excitation rates for astrophysical abundant
ions with open-shell configurations $n{\rm p}$,  $n{\rm p}^5$,
$n{\rm p}^2$ and $n{\rm p}^4$ which give rise to infrared (IR)
lines (see Figure~\ref{IR_lines}). The following isoelectronic
sequences were studied: for $n=2$, the B-, F-, C- and O-like
ions; and for $n=3$, the Al-, Cl-, Si- and S-like ions. These
data were required for the modeling of spectra observed with
the Infrared Space Observatory (ISO\footnote{\tt
http://iso.esac.esa.int/}) launched in November 1995, and after
August 2003, with the Spitzer Space Telescope
(Spitzer\footnote{\tt http://spitzer.caltech.edu/}). A delicate
situation emerged in the electron excitation of IR transitions
which is illustrated \cite{but94} with the ${\rm 2s}^2{\rm
2p}^4\, ^3{\rm P_0}\rightarrow\, ^3{\rm P_1}$ transition in
O-like Al~VI (see Figure~\ref{butler_zeippen}). For high-charge
species (e.g. Al~VI), a bundle of broad resonances that belong
to series converging to the levels of the first excited
configuration, ${\rm 2s}{\rm 2p}^5$, sits just above the
reaction threshold. As a consequence, the effective collision
strength at the lower temperatures is not only greatly enhanced
but also becomes very sensitive to target-level separations
\cite{sar94} (see Figure~\ref{saraph_tully}); i.e. experimental
target levels must be used in the $R$-matrix runs and the
convergence of the close-coupling expansion (\ref{cc}) with
respect to target levels is crucial. An extreme case
\cite{sto96} is the ${\rm 3s}^2{\rm 3p}\ ^2{\rm
P^o_{1/2}}\rightarrow\, ^2{\rm P^o_{3/2}}$ transition in
Al-like Fe~XIV which, as shown in Figure~\ref{storey_etal},
displays a dense resonance structure arising from excited
levels from the $n=3$ and $n=4$ complexes. An important feature
here is the choice of energy mesh interval which must be fine
enough to resolve the resonances; in the region between the
${\rm 3s}^2{\rm 3p}\ ^2{\rm P^o}$ and ${\rm 3s3p}^2\ ^4{\rm P}$
thresholds, an energy mesh with 10,500 points was used
\cite{sto96}. Finally, a compilation \cite{bad06} of IP
$A$-values and effective collision strengths for interpreting
IR transitions in nebulae was published in 2006.


\begin{figure}
\centering
\caption{\label{saraph_tully} Effective collision strength for the
${\rm 2s}^2{\rm 2p}^5\ ^2{\rm P^o_{3/2}}\rightarrow\, ^2{\rm P^o_{1/2}}$
transition in F-like Ar~X (a) using theoretical level separation and (b)
using experimental level separation. Differences around a factor of 2
are observed at $\log T = 4$. Reproduced from Figure~5 of
Ref.~\cite{sar94} (http://adsabs.harvard.edu/abs/1994A\%26AS..107...29S).}
\end{figure}


\begin{figure}
\centering
\caption{\label{storey_etal} Effective collision strength for the
${\rm 3s}^2{\rm 3p}\ ^2{\rm P^o_{1/2}}\rightarrow\, ^2{\rm P^o_{3/2}}$
transition in Al-like Fe~XIV in the region below the ${\rm 3s3p}^2\ ^4{\rm P}$
excited levels showing a dense resonance structure. Reproduced from
Figure~2 of Ref.~\cite{sto96} (http://adsabs.harvard.edu/abs/1996A\%26A...309..677S).}
\end{figure}


The second stage of the IP was concerned with the computation
of electron impact excitation rates for transitions within the
$n\leq 3$ complexes and, when possible, $n\leq 4$ in ionic
species of the complete Fe isonuclear sequence. These data are
mainly needed in the interpretation of the EUV solar spectrum,
and proved to be quite an ambitious objective because for some
systems
--- e.g. second-row ions with open $n=3$ shells such as the
Al-, Si-, P-, S-, Cl-like species
--- the calculations proved to be very large indeed due to the
slow convergence of the close-coupling expansion (\ref{cc})
with respect to target levels and the need to include
relativistic effects. In fact, some of them, e.g. the Fe~XI
(S-like) and Fe~XIII (Si-like), have not as yet been published.

A topical example of this second stage is the electron impact
excitation of the ground state of Al-like Fe~XIV. In this
calculation \cite{sto00}, the target representation included 18
$LS$ terms arising from the configurations ${\rm 3s}^2{\rm
3p}$, ${\rm 3s}{\rm 3p}^2$, ${\rm 3s}^2{\rm 3d}$, ${\rm 3p}^3$
and ${\rm 3s}{\rm 3p}{\rm 3d}$ with the addition of 17
correlation configurations containing $nl$-orbitals with $n\leq
4$. Intermediate-coupling collision strengths were obtained
with the algebraic transformation of Eq.~(\ref{jajom})
including relativistic target effects via term-coupling
coefficients. Previous work \cite{duf91} was carried out with
the close-coupling method ($R$-matrix) using a reduced
three-configuration target representation, and in the distorted
wave approximation \cite{bha93, bha94} which neglects channel
coupling. In this respect, it might be seen in
Figure~\ref{storey_etal2} that, although the effective
collision strengths for the ${\rm 3s}^2{\rm 3p}\ ^2{\rm
P^o_{1/2}}\rightarrow {\rm 3s}{\rm 3p}^2\ ^2{\rm D_{5/2}}$
transition computed with the three methods agree at high
temperatures ($\log T\approx 7$), at the lower temperatures
($\log T\approx 5.5$) the IP is a factor of 2 higher; this
enhancement is caused by an adequate treatment of resonances by
the IP. Furthermore, as depicted in Figure~\ref{storey_etal3}
and in contrast with IP, previous work resulted in a density
insensitive 274$\lambda$/270$\lambda$ line ratio which was
astronomically observed to vary between 1.3 and 2.3. The IP
variation was found to be between 1.0 and 2.0 in good agreement
with observations.


\begin{figure}
\centering
\caption{\label{storey_etal2} Effective collision strength for the
${\rm 3s}^2{\rm 3p}\ ^2{\rm P^o_{1/2}}\rightarrow
{\rm 3s}{\rm 3p}^2\ ^2{\rm D_{5/2}}$ transition in Al-like Fe~XIV.
Solid line: IP close-coupling calculation \cite{sto00}. Open circles:
independent close-coupling calculation \cite{duf91} with a reduced
target representation. Dashed line: distorted wave calculation
\cite{bha93} that neglects channel coupling. Reproduced from
Figure~2 of Ref.~\cite{sto00} (http://dx.doi.org/10.1051/aas:2000319).}
\end{figure}


\begin{figure}
\centering
\caption{\label{storey_etal3} Density variation of the
274$\lambda$/270$\lambda$ line ratio in Fe~XIV. Solid line: IP work
\cite{sto00}. Dashed line: CHIANTI~v1.0 using previously computed
\cite{duf91} close-coupling effective collision strengths. Asterisks:
distorted wave data \cite{bha93, bha94}. Reproduced from Figure~4
of Ref.~\cite{sto00} (http://dx.doi.org/10.1051/aas:2000319).}
\end{figure}


It may be assumed that resonances enhance the effective
collision strength mainly at the lower temperatures but this is
not always so. In such cases, a detailed study of the whole
resonance structure and the inclusion of all the contributing
thresholds in the target representation are required. This
problem is illustrated \cite{ber97} with the excitation  of the
${\rm 2p_{1/2}}\rightarrow {\rm 2p_{3/2}}$ transition in
Li-like Fe~XXIV (see Figure~\ref{berrington_tully}) where a
huge bump with respect to a distorted wave result \cite{zha90}
occurs at around $\log T\approx 7$. Moreover, estimating the
right number of contributing thresholds for the ${\rm 3s}^2{\rm
3p}^q$ ions ($q=2{-}4$) has become --- as experienced
\cite{bin01, bin98a, bin98b, sto05} with Fe~XII ($q=3$) --- a
long ordeal; however, as targets were improved, long-standing
discrepancies with observations were progressively ironed out
\cite{bin01, sto05}.


\begin{figure}
\centering
\caption{\label{berrington_tully} Effective collision strength
for the  ${\rm 2p_{1/2}}\rightarrow {\rm 2p_{3/2}}$ transition in
Li-like Fe~XXIV. Full line: IP \cite{ber97}. Broken line: distorted
wave result \cite{zha90}. The IP data show a huge bump at
$\log T\approx 7$ due to the resonance contribution. Reproduced
from Figure~9 of Ref.~\cite{ber97} (http://dx.doi.org/10.1051/aas:1997384).}
\end{figure}

An interesting calculation \cite{pel97} has been the low energy
electron impact excitation of neutral iron (Fe~I) since most of
the IP work involved positively charged targets. An important
effect here was to take into account the polarizability of the
terms included in the close-coupling expansion, namely ${\rm
^5D}$, ${\rm ^5F}$ and ${\rm ^3F}$, by considering (at least
for the first two) four polarized pseudo-states ($\overline{\rm
^5P^ o}$, $\overline{\rm ^5D^ o}$, $\overline{\rm ^5F^ o}$ and
$\overline{\rm ^5G^ o}$). A second challenge was the threshold
behavior of the collision strength for a neutral,
$\Omega(E)\rightarrow 0$ as $E\rightarrow 0$, which differs
from that for an ionic target ($\Omega(E)\rightarrow
\Omega_{\rm th}$ as $E\rightarrow 0$), and obeys the Wigner
threshold law
\begin{equation}
\Omega(i,f)\sim E^{l+1/2}
\end{equation}
where $l$ is the dominant orbital angular momentum
contribution. Compliance with this law may be appreciated for
the case of the $^5{\rm D_4}\rightarrow\, ^5{\rm D_3}$
transition shown in Figure~\ref{pelan_berrington}, where at
very low energies ($E\lesssim 5\times 10^{-3}$~Ryd), the s-wave
($l=0$) dominates while at higher energies the p-wave ($l=1$)
takes over.


\begin{figure}
\centering
\caption{\label{pelan_berrington} Low-energy behavior of the
collision strength  $\Omega(^5{\rm D_4}, ^5{\rm D_3})$
in Fe~I showing compliance with the Wigner threshold law. Full curve:
IP close-coupling calculation \cite{pel97}. Dotted line: Wigner law
for s-wave ($l=0$). Dashed line: Wigner law for p-wave ($l=1$).
Reproduced from Figure~1 of Ref.~\cite{pel97} (http://dx.doi.org/10.1051/aas:1997328).}
\end{figure}


Massive data sets of relativistic radiative data for both
allowed and forbidden bound--bound transitions were also
computed in the IP for several Fe ions (e.g. Fe~XVII in
Ref.~\cite{nah03}), increasing previously available data sets
for such ions by orders of magnitude. For allowed transitions,
the BPRM package was used where level identification became an
issue, greatly simplified by the development of a new level
identification algorithm. For the forbidden type (E2, E3, M1
and M2), the SUPERSTRUCTURE atomic structure code was employed.
Photoionization cross sections for bound--free transitions were
also computed with BPRM for some species, an oustanding case
\cite{bau97} was Fe~I for which cross sections were computed
for 1,117 levels with $n\leq 10$ and $l\leq 7$. The Fe~II
target representation in this large calculation required 52
$LS$ terms from the configurations ${\rm 3d}^6{\rm 4s}$, ${\rm
3d}^7$, ${\rm 3d}^6{\rm 4p}$, ${\rm 3d}^5{\rm 4s}^2$ and ${\rm
3d}^5{\rm 4s4p}$. In Figure~\ref{bautista} the photoionization
cross section of the ground state of Fe~I from this work is
compared with a previous theoretical estimate \cite{kel72}.


\begin{figure}
\centering
\caption{\label{bautista} Photoionization cross section of the
${\rm 3d}^6{\rm 4s}^2\ ^5{\rm D}$ ground state of Fe~I where the
contributing thresholds are denoted. Full curve: IP close-coupling
result \cite{bau97}. Broken curve: calculation \cite{kel72} based
on many-body perturbation theory. Reproduced from Figure~2 of
Ref.~\cite{bau97} (http://dx.doi.org/10.1051/aas:1997327).}
\end{figure}


\section*{TIPbase}

The TIPbase\footnote{\tt
http://cdsweb.u-strasbg.fr/tipbase/home.html} database contains
the fine-structure atomic data computed in the IP, namely level
energies, radiative transition probabilities ($A$-values) and
electron impact excitation cross sections (collision strengths)
and rates (effective collision strengths) for fine-structure
transitions. Its DBMS was developed in fortran 90 and its web
user interface in javascript, the latter by Jes\'us Quiroz
(IVIC) and Marcio Mel\'endez (Universidad Sim\'on
Bol\'{\i}var). Among its most interesting features are
extensive data documentation and the possibility to display and
manipulate plots interactively with a java applet, which may be
used to study trends and resonances in the excitation cross
sections. Efforts have been made to include the bulky data sets
containing collision strengths, which allow the computation of
rates for non-Maxwellian distributions, photoionization cross
sections and recombination rates. The inclusion of the latter
parameter was phased out, however, due to differences in the IP
on how to compute recombination rates. Up to the present,
TIPbase contains only IP data which perhaps limits its
usability as it mainly concentrates on the iron isonuclear
sequence.


\section*{CHIANTI}

CHIANTI\footnote{\tt http://www.chianti.rl.ac.uk/} is an atomic
data application for spectroscopic diagnostics of astrophysical
plasmas \cite{der97}, maintained since the end of the 90s by an
international collaboration that includes several IP members.
As described in its six releases\footnote{\tt
http://www.chianti.rl.ac.uk/publications/pub.html}, it may be
regarded as a computer tool --- developed with a friendly
user-interface based on the Interactive Data Language (IDL) ---
to calculate synthetic, optically thin, emission-line spectra
of astrophysical sources and to derive plasma diagnostics. It
makes use of an extensive and periodically assessed and
upgraded atomic database which caters mainly for the solar
community and, more recently, for X-ray work. For a large
number of ions, its database contains energy levels,
wavelengths, radiative transition probabilities, electron
impact excitation rates (many from the IP) and the continuum.
More recently, CHIANTI has been overhauled with inner-shell
transitions, satellite lines, proton excitation, two-photon,
relativistic free--free and free--bound continua, ionization
and recombination rates and Fe K lines. The impressive CHIANTI
citation list\footnote{\tt
http://www.chianti.rl.ac.uk/publications/citation\_list.html}
certainly gives an idea of the extent of its current user base.


\section*{Current projects}

Although the OP and IP are still on-going enterprises, most of
their members are involved in other atomic data computational
projects. Some (e.g. the RmaX Network) may be regarded as
spin-offs from the OP/IP, but others, although completely
independent, carry the Seaton spirit and the expertise acquired
in the past two decades of team work. We briefly mention here a
few of the more active collaborations.


\subsection*{The UK APAP/RmaX Network}

The Atomic Processes for Astrophysical Plasmas Network of the
United Kingdom (APAP\footnote{\tt
http://amdpp.phys.strath.ac.uk/UK\_RmaX/}) is a continuation of
the original UK\_RmaX Network which was a spin-off of the IP.
It is dedicated to the calculation of radiative and collisional
data for the modeling of astrophysical plasmas and spectral
analysis. Data sets containing both fundamental and derived
atomic data are made available, and much work has been done in
the development and upkeep of the $R$-matrix codes and atomic
structure codes, particularly in relation to improvements in
access, user interfaces ({\tt NAMELIST} based) and
parallelization. Recent publications\footnote{\tt
http://amdpp.phys.strath.ac.uk/UK\_RmaX/APAP\_pub.html}
concentrate on topics such as atomic-data benchmarking,
inner-shell electron impact excitation, radiative and Auger
damping, fluorescence yields, electron--ion recombination,
dielectronic recombination and EUV line identifications and
diagnostics.


\subsection*{Work at OSU}

Recent work at the Ohio State University (OSU), one of the most
active members of the IP, has concentrated on contributions to
the RmaX Network and in the computation of electron--ion
recombination rate coefficients and photoionization cross
sections for astrophysically abundant elements (e.g. Fe~XVI,
Fe~XVII and Fe~XXI). An interesting new approach has been the
relativistic coupled-cluster method used \cite{sur08} to
compute the transition probability of K$\alpha$ lines in F-like
ions ($Z=10{-}79$), finding a K$\alpha_1$ and K$\alpha_2$ cross
over at $Z=41{-}42$. In the extensive work on recombination
(see Ref.~\cite{nah06} and references therein), high-precision
photoionization and recombination cross sections and rates are
computed self-consistently with BPRM using the same
wave-function expansion, thus treating radiative and
dielectronic recombinations in a unified manner \cite{nah04}.
It is then shown \cite{nah06b} that resonances in the unified
recombination cross sections correspond directly to
dielectronic satellite spectra. Moreover, in the context of the
dielectronic satellite lines of He-like Fe~XXV and Ni~XXVII,
this method has been recently compared \cite{nah09} with the
isolated resonance approximation commonly used in plasma
modeling; good agreement (20\%) is obtained for the rates of
strong lines, but larger discrepancies are found for the weaker
features. Large data sets containing $A$-values and $gf$-values
for many ions are being computed using BPRM for fine-structure
E1 transitions and SUPERSTRUCTURE for the forbidden type. All
radiative data from OSU are available online from the
NORAD-Atomic-Data\footnote{\tt
http://www.astronomy.ohio-state.edu/$\sim$nahar/nahar\_radiativeatomicdata/index.html}
web site.


\subsection*{xstarDB}

XSTAR \cite{bau01} is a widely used spectrum modeling package
that computes the physical conditions and emission spectra,
particularly in the X-ray region, of a photoionized gas. It can
be run either as a single model or as a grid of models, the
latter to estimate model sensitivity to input-parameter ranges.
For the past eight years, a considerable effort has been
focused on improving its atomic database (xstarDB) with
accurate inner-shell radiative and collisional data (K-vacancy
level energies, wavelengths, $A$-values, Auger widths,
photoabsorption cross sections and electron impact excitation
cross sections) which are required in the analysis of K lines,
particularly from nitrogen \cite{gar09}, oxygen \cite{gar05},
second-row elements (Ne, Mg, Si, S, Ar and Ca) \cite{pal08a,
wit09}, iron \cite{pal02, bau03, pal03a, pal03b, men04, bau04}
and nickel \cite{pal08b}. This is an interesting project in as
much as both data producers and users are working closely
together to generate a tailor-made atomic database.

Among the most relevant findings in this work \cite{pal02} has
been the smearing  of the K edge caused by both radiation and
Auger dampings, exemplified with the K photoabsorption of the
Fe~XVII ground state which is dominated by the double series of
${\rm 1s2s}^2{\rm 2p}^6n{\rm p}\ ^{3,1}{\rm P^o_1}$ resonances
(see Figure~\ref{palmeri_etal}). These resonances decay via the
following manifold:

\begin{eqnarray}
{\rm 1s2s}^2{\rm 2p}^6n{\rm p} & \stackrel{{\rm K}n}{\longrightarrow}
         \label{beta}       & {\rm 1s}^2{\rm 2s}^2{\rm 2p}^6+h\nu_n \\
         \label{alpha}       & \stackrel{{\rm K}\alpha}{\longrightarrow} &
                  {\rm 1s}^2{\rm 2s}^2{\rm 2p}^5n{\rm p}+h\nu_\alpha \\
         \label{part}     & \stackrel{{\rm KL}n}{\longrightarrow} &
                          \begin{cases}
                  {\rm 1s}^2{\rm 2s}^2{\rm 2p}^5+e^- \\
                  {\rm 1s}^2{\rm 2s}{\rm 2p}^6+e^-
                          \end{cases}  \\
         \label{spec}     & \stackrel{{\rm KLL}}{\longrightarrow} &
                          \begin{cases}
                  {\rm 1s}^2{\rm 2s}^2{\rm 2p}^4n{\rm p}+e^- \\
                  {\rm 1s}^2{\rm 2s}{\rm 2p}^5n{\rm p}+e^- \\
                  {\rm 1s}^2{\rm 2p}^6n{\rm p}+e^-
                          \end{cases}
\end{eqnarray}
which is distinctively dominated by the radiative K$\alpha$
(Eq.~\ref{alpha}) and Auger KLL (Eq.~\ref{spec}) channels where
the active $n{\rm p}$ electron remains a spectator; therefore,
their decay transition probabilities are independent of $n$.
Consequently, as seen in Figure~\ref{palmeri_etal}, if the
K$\alpha$ and Auger KLL decay channels are not taken into
account, i.e. damping is neglected, the resonances are narrow
and asymmetric and converge towards a sharp K edge. On the
other hand, if damping is included, the resonance series
display constant widths and symmetric profiles that become
progressively smeared with increasing $n$ to produce a smooth
transition through the K threshold. Such edge smearing may be
of relevance to the astrophysical interpretation \cite{kal04}
of absorption features in astrophysical X-ray spectra.


\begin{figure}
\centering
\caption{\label{palmeri_etal} Total photoabsorption cross
section of the ground state of Ne-like Fe~XVII in the region
near the K edge (7.72 keV). (a) Damping has been neglected.
(b) Radiation damping is included. (c) Radiation and spectator
Auger dampings are included. Reproduced from Figure~1 of
Ref.~\cite{pal02} (http://dx.doi.org/10.1086/344243).}
\end{figure}


\subsection*{ADAS-EU}

The Atomic Data and Analysis Structure for Fusion in Europe
(ADAS-EU\footnote{\tt http://www.adas-fusion.eu/}) is a project
that enables plasma diagnostics and modeling for fusion
laboratories in Europe and, in the near future, for
ITER\footnote{\tt http://www.iter.org/}. Databases containing
both fundamental and derived atomic data are implemented and
maintained, also promoting the computation and measurement of
new data. It has grown from the active ADAS
consortium\footnote{\tt
http://www.adas.ac.uk/documentation.php} which developed a set
of computer codes, subroutine libraries and atomic data sets
for plasma modeling and the analysis and interpretation of
spectral emission. New topics of interest are: heavy element
spectroscopy and models; charge exchange spectroscopy; beam
stopping and beam emission spectroscopy; diatomic spectra and
collisional--radiative models.


\section*{E-science and the data deluge}

With increasing frequency and effectiveness, researchers are
exchanging data, ideas, publications, references and images
through a variety of electronic means. However, a number of
innovative collaborative environments are emerging which have
not been readily adopted by the science community as its
everyday workplace \cite{bor06, col06, col07, der07}. This new
ubiquitous mode of collaborations is being mainly supported by
e-mail and instant messaging. Beyond this exchange among peers,
the \textit{post-Gutenberg} era has dawned where information
producers, i.e. researchers, research centers and academic
institutions, now have the ability to publish and disseminate
their intellectual production without intermediaries.

Terms such as ``cyber-infrastructure'', ``e-science'' and, more
recently, ``e-research'' have been coined to describe this
knowledge revolution. Among its most distinctive features we
can cite: the intensive use of information and communication
technologies (ICT); geographically distributed resources for
information processing and analysis; and above all, its
ubiquitousness (see Refs. \cite{fos05, hey03, hey05} and
references therein). Its main challenge is to manage, analyze
and preserve the \textit{data deluge} caused by an impressive
variety of sensors, ambitious numerical simulations and
large-scale facilities such as particle accelerators, tokamaks,
synchrotron radiation sources and ground and satellite-borne
astronomical telescopes. Under this digital avalanche, which
surpasses any traditional data management capacity, ICT can
transform these instruments into powerful computer environments
for data mining.

As we have illustrated here with the OP/IP, large data
productions are usually carried out by global collaborations,
i.e. multinational science groups that generate large volumes
of data which are geographically distributed and maintained
only during project life cycle. Most of these data are never
published and, when the collaborations end, many are lost or
stashed away in national (or international) reservoirs that
have nothing to do with their origins. Production decisions,
approximations and provenance are buried in a huge electronic
correspondence to which no-one has access \cite{gra02}.

A similar path is followed by small data producers scattered
around the globe; thus both large and small data producers face
the same problems in knowledge cataloging, preservation and
dissemination. It is imperative to plan and build repositories
that store data as they emerge and to retain the history of the
decisions and criteria that generate them \cite{bor07, gra02,
kar06, mur08}. In spite of pioneering efforts more than a
decade ago to create a framework of recommendations to guide
scientific database preservation and dissemination
\cite{doz95}, it is only recently that multilateral
organizations and planners in Europe and the United States have
started to generate technical reports to encourage the
preservation of important scientific data collections
\cite{arz04a, cla07, lor03, lyo07, pil07, sim05, dph09}.
However, many of these recommendations have not permeated to
the producing communities and/or to the collection custodians
in these countries. The situation is even worse in Latin
America where we are still not convinced by, or at least aware
of, the new paradigms in the production and dissemination of
scientific knowledge, and consequently, only a low-level use of
ICT has been incorporated in teaching and research.

This increasing awareness to preserve and share data for use
and reuse in knowledge creation has given the Open Access
Movement a new facet. The self-preservation and dissemination
of scientific data are beginning to be considered and discussed
in different scenarios \cite{arz04b, cla07, lyo07, pil07,
sab07, dph09, xu07}. Inspired by the reflections and conceptual
bases of the debutante, the Open Data Movement, it is essential
to establish standards and protocols such that research data
can be processed without barriers or high costs. The academic
community perceives published data as their heritage; however,
many publishers are insisting on copyright which, without
doubt, is one of the greatest obstacles in e-research
\cite{mur08}.


\section*{Data \& metadata in e-science}

The reasons for preserving data derive from the fact that
observations, knowledge and understanding are cumulative. A
{\em datum} can be considered a piece of information that can
be processed, interpreted, transmitted and preserved. Data
arise either from measurements (observational or experimental
data), simulated results (synthetic data computed with
mathematical models) or historical records (historical data).
Data may be considered raw if they are generated directly from
measurements and models, or derived if they undergo further
filtering and processing. But, additionally, there are many
subtleties and complexities: what might be considered derived
data for some processes are raw for others, and historical data
may emerge from the mathematical models that produce the
record.

E-research requires the automatic handling of large volumes of
data which must be streamlined by standards specified by both
the data producers and users. The basic information employed to
describe data --- its content, features, dates, terms of use,
source, ownership and other characteristics --- is referred to
as {\em metadata}. The latter allows the user to evaluate
whether a particular data set is suitable for their purposes
and to facilitate information access. Adequate documentation
regarding sampling, analytical procedures, anomalies, accuracy
and structure is of vital importance in its future correct
interpretation. Metadata can facilitate \cite{mic97}:
\begin{itemize}
  \item data identification and acquisition for a given
      subject, for a specific period of time or geographic
      location;

  \item automatic analysis and data modeling;

  \item the inclusion of semantic knowledge elements
      associated with the data.
\end{itemize}

Metadata standardization is important inasmuch as it allows the
definition of common terminologies specifying entry,
validation, access, integration and synthesis in automation,
and ensures complete and accurate documentation of data-set
content. There are different metadata standards available,
namely Dublin Core, Darwin Core, Content Standard for Digital
Geospatial Metadata, ISO 19115 Geographic information metadata,
Ecological Metadata Language, etc. The reason for so many
standards is the diversity of application fields: information
science, biology, geology, ecology and cartography, to name but
a few.

Structures (classes) are composed of metadata items associated
with descriptive semantic definitions for some of the possible
data attributes. Such structures can be arbitrarily simple or
very complex, and the information they contain can be
heterogenous depending on the data types and needs in the
associated communities. Within its metadata model, each
community can define a property or an item differently; for
instance, the Dublin Core initiative specifies a base set of 15
elements while the metadata model of the Learning Objects
Metadata, which is being developed by the IEEE and other
organizations to describe resources in teaching--learning
environments, has about 100 items \cite{gue04, tor03}.

The incorporation of metadata demands an investment of time and
effort by those who generate, preserve and share the data. It
is advisable to make allocation for metadata-model definition
and for the implementation learning curve, followed by
maintenance costs in the short, medium and long term. For the
implementation of a metadata system to be successful, there
must be institutional commitment, i.e. the acceptance of
technical field staff, researchers, students and computer and
laboratory technicians.

All this effort takes place naturally in big-science
experiments; however, it is also necessary for research
consortia and groups and individual investigators to become
aware of the importance of cataloging and preserving data. Only
by doing so will their relevance to future generations be
ensured \cite{bor07}.


\section*{Scientific Communities and DDCEs}

A network of digital data curation environments (DDCEs) is
beginning to be established by communities pursuing scientific
knowledge for the consolidation and online analysis of the
digital avalanches produced by their instruments. Digital
curation involves the structuring and maintenance of such data
for the use of current and future generations (see Refs.
\cite{bea06, bea07} and references therein). A DDCE is then a
set of services and tools committed to capturing, preserving,
curating and disseminating data. From the information point of
view, a DDCE tends to be cumulative, permanently open
--- both in content and in the platform that supports it
\cite{gra02b, lor03, lor04} --- and plays a key role in
preserving and replicating the community's memory. In fact, a
DDCR is more than simply a storage of data objects and a
toolbox for data mining and analysis, as researchers will need
to interact with data within the environment of a social
network and with the Web 2.0 approach. This new and challenging
reality therefore introduces significantly greater complexity
to data management \cite{luc08}.

Some progress has been made in developing simulations and data
capture tools, but not much on those for data analysis. It is
common to  build new instruments with little planning or
budgeting provision for data management \cite{gra09}. Much less
can be said about social software with collaboration
capabilities which is the emerging paradigm for scientific
activities. What we really need are {\em workflow} environments
to set up, with Web 2.0 functionalities, a pipeline from the
instrument or simulation directly to the DDCE (see articles in
Ref.~\cite{hey09} and references therein).

A DDCE must be compliant with the Open Archive Initiative
(OAI); that is, the data repositories should be available on
the Internet, allowing any user to read, download, distribute,
print, search and link the data files, and to use them for any
other lawful purpose without financial, legal or technical
barriers other than those associated with Internet access
itself \cite{ mur08, xu07}.

In the next sections we describe some of the efforts that have
been made by different communities to curate, disseminate and
preserve their digital patrimony.


\subsubsection*{CombeChem \& e-Bank: a proof of concept}

One of the first experiences in building a DDCE is taking place
at the UK National Crystallography Service (NCS). This service
has developed, in collaboration with the CombeChem\footnote{\tt
http://www.combechem.org/} e-science testbed and the eBank-UK
project, an e-infrastructure to facilitate end-to-end a
crystallographic experiment. A proof of concept that integrates
existing structure and proprietary data sources has been
developed within a grid-based information- and
knowledge-sharing environment \cite{col05, col06, duk05}.

Following the OAI concept, these projects have constructed a
data repository, eCrystals\footnote{\tt
http://ecrystals.chem.soton.ac.uk/}, that makes the raw and
derived data from a crystallographic experiment available. The
data are uploaded into a repository where they are complemented
with additional metadata (chemical and bibliographic). This
approach allows the rapid release of crystal structure data
into the public domain, but also provides mechanisms for value
added services that foster data identification for further
studies even though the ownership of the data is always
retained by the producer. Publication of all the results
generated during the course of the experiment is then enabled
by means of an Open Access Data Repository.

The end product of this project is a growing number of
participating partners managing data repositories. There are
practical implications that reflect the changing nature of
research practices towards the data-intensive paradigm, with a
variety of workflows in smart laboratories where scientists
require more tools and services for virtual experiments. The
emerging eCrystals Federation is thus gathering useful
experience in creating a network of institutional ePrint
repositories \cite{lyo08}.


\subsubsection*{Astronomy \& Astrophysics: a leading e-data community}

Astronomical data are growing at an exponential rate from the
continuous construction of new telescopes with ever more
sensitive detectors and from ambitious computational models.
While instruments produce steady data streams, there is an
increasingly complex worldwide network of facilities with large
data outflows. The natural characteristics of the astronomical
community have made it an early builder and adopter of DDCEs,
the most relevant being: a unified taxonomy, vocabulary and
coded definition of metrics and units; peer reviewed data
carefully collected with rigorous statistical standards;
trackable data provenance; publicly available data; and ancient
data preservation as they are essential for time-varying
phenomena \cite{gra02b}.

For years this community has been building a knowledge platform
that has become a revolution in the way astronomers use data,
providing a role model to other disciplines on how technology
can be used to improve the quality and effectiveness of
scientific enquiry \cite{nor06}. These unique data services
involve: three important data centers --- the SAO/NASA
Astrophysics Data System (ADS\footnote{\tt
http://www.adsabs.harvard.edu/}), the Centre de Donn\'ees
astronomiques de Strasbourg (CDS\footnote{\tt
http://cdsweb.u-strasbg.fr/}) and the NASA/IPAC Extragalactic
Database (NED\footnote{\tt http://nedwww.ipac.caltech.edu/})
---; the International Virtual Observatory Alliance
(IVOA\footnote{\tt http://www.ivoa.net/}) \cite{djo05}; the
public data releases from individual astronomical projects; the
preprint dissemination from the {\em arXiv/astro-ph} server;
and the rapid dissemination of results made possible by
forward-thinking journals. Most of the data repositories are
open-access compliant and enthusiastically promote the Open
Data Movement. In addition to the data services, the IVOA
offers the community a toolkit of online software for data
analysis. Furthermore, the astronomical community is sensitive
to the ``digital divide", thus allowing open access to
astronomers in developing countries.

This leading community is now facing new kinds of difficulties
to be considered by the other emerging knowledge groups,
despite information on the variety of resources available.
International pressure against the open access of astronomical
data is appearing, and it seems there is not enough awareness
to counteract it. Difficulties remain between journals and data
centers, and most of the published data are never posted by the
latter. New instruments are still being built with little
planning or budgeting for data management, and there is little
provision for digital preservation. \cite{nor05, nor06}.


\subsubsection*{ITER: an emerging community}

The International Thermonuclear Experimental Reactor
(ITER\footnote{\tt http://www.iter.org}) is an international
large-scale tokamak experiment aiming to demonstrate the
possibility of producing commercial energy from magnetic
confinement fusion, expecting to do so about a decade from now.
One of the greatest challenges for the ITER systems will be the
capability for analyzing, visualizing and assimilating the data
to support decision making, and therefore, the scientific
productivity of ITER will be inextricably linked to the power
of its collaborative infrastructure. The need to support remote
operation of the experimental facilities will require
collaborative efforts involving over 2,000 scientists
worldwide. The fusion community is moving towards the vision of
remote hardware control of the experiment \cite{gre05, sch08,
sch09}, and is aware of the importance of data curation before
the launch of the experiment; e.g. the development of a set of
software tools (MDSplus\footnote{\tt http://www.mdsplus.org}
for Model Data System) for data acquisition and storage. In
fact, it is more than that as it pretends to be a methodology
for the management of complex scientific data. Furthermore, it
is worth mentioning the effort for implementing an atomic
database infrastructure throughout the project, namely the
``Atomic Data and Analysis Structure for Fusion in Europe"
(ADAS-EU\footnote{\tt http://www.adas-fusion.eu/}). This
project aims to improve the effectiveness of data analysis in
existing fusion experiments and prepare for ITER.


\section*{VAMDC}

The Virtual Atomic and Molecular Data Centre
(VAMDC\footnote{\tt http://www.vamdc.org/}) is a multinational
project launched in July 2009 involving 24 research teams from
several countries of the European Union (Austria, France,
Germany, Italy, Sweden and the United Kingdom), Russia, Serbia
and Venezuela. The Venezuelan node involves CeCalCULA and the
IVIC Computational Physics Laboratory. Atomic and molecular
data (A\&M) are required in a wide variety of scientific
fields, e.g. astrophysics, fusion plasmas, atmospheric physics
and chemistry, environmental studies and quantum optics, and
technological applications such as lighting, semiconductors,
nanotechnology and molecular biology. The VAMDC intends to
upgrade and integrate at least 21 A\&M databases
--- TOPbase, TIPbase, OPserver, xstarDB and CHIANTI among them
--- in order to implement an interoperable cyber-infrastructure
for the exchange of atomic and molecular data. Its principal
investigator is Marie-Lise Dubernet (Universit\'e Pierre et
Marie Curie and Observatoire de Paris).

The partners, roadmap and basic guidelines of the VAMDC have
been previously described \cite{men09}, but the following
dimensions can be emphasized.
\begin{itemize}

\item {\bf Network activities.} They are aimed at
    coordinating infrastructure activities between
    interdisciplinary fields and other relevant projects
    such as Astrogrid\footnote{\tt
    http://www.astrogrid.org/}, ITER\footnote{\tt
    http://www.iter.org/} and Europlanet\footnote{\tt
    http://www.europlanet-eu.org/}. They are to promote the
    VAMDC services and installations involving data users
    from other disciplines such as astrophysics,
    atmospherics and fusion.	

\item {\bf  Services.} The VAMDC is establishing a
    cyber-infrastructure for A\&M data producers and users,
    giving access to the main databases in an interoperable
    format, maintaining registries, dictionaries and demand
    nodes, and implementing distributed grid environments
    for applications and data reservoirs. 	

\item {\bf Joint research.} Interdisciplinary teams are
    developing computational tools to sustain the e-science
    platform of the VAMDC, defining standards, protocols
    and specifications for the storage, exchange,
    publication and mining of A\&M data.

\end{itemize}


\begin{figure}
\centering
\includegraphics[width=10cm]{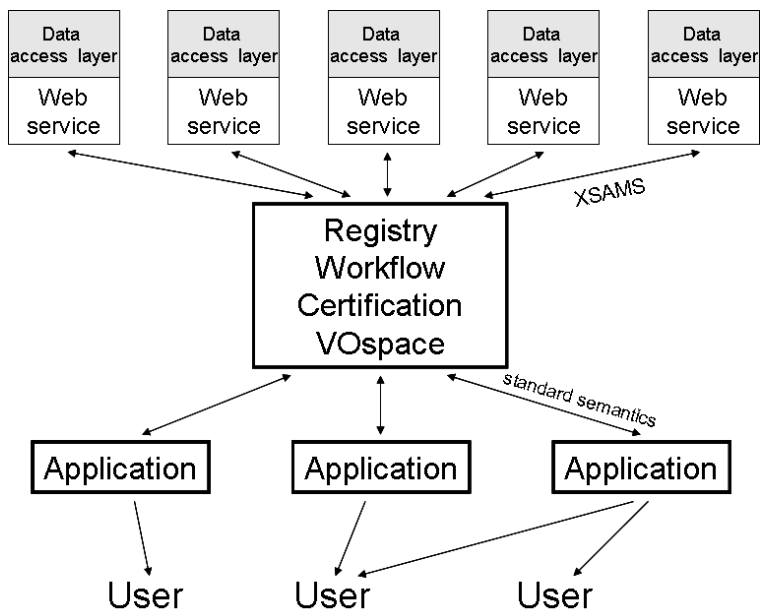}
\caption{\label{vamdc} VAMDC virtual data-warehouse distributed structure.}
\end{figure}


The VAMDC is addressing some of the chronic problems in A\&M
data activities \cite{men00}. For instance, due to the lack of
funds, standards and common guidelines, outstanding problems in
existing A\&M database are interoperability and data interfaces
which hamper productive searches and data mining. Data exchange
has also been carried out in a somewhat informal manner
(emails, undocumented ASCII files, peer-to-peer arrangements)
even though standard formats from specific client disciplines
(e.g. the FITS\footnote{\tt
http://heasarc.gsfc.nasa.gov/docs/heasarc/fits\_overview.html}
astronomical data format) have been incorporated. In this
respect, one of the initial VAMDC policies has been the release
of an XML schema, referred to as XSAMS\footnote{\tt
http://www-amdis.iaea.org/xsams/about.html}, for the exchange
of A\&M data which is at present under evaluation. Moreover,
A\&M data are stored in a variety of relational DBMSs using
very diverse data models. Many of these DBMSs are not SQL
standard packages but local developments that, in the long run,
can compromise data integrity and regular updating procedures.
In some cases data are not even housed in databases but as
plain files in data centers, servers belonging to scientific
journals or in departmental, project and personal web pages.
Therefore, XSAMS becomes the key not only for data exchange but
also for data identification and provenance when implementing a
new generation of search engines.

In Figure~\ref{vamdc} we show the structure of the VAMDC which
may be visualized as a virtual data-warehouse addressing a
collection of heterogenous, distributed web services. To
facilitate end-user data mining, the warehouse contains
registries, dictionaries, catalogs, workflows, metadata, a
middleware and a uniform XML-based format for data transfers
(XSAMS).


\section*{OP/IP perspectives within VAMDC}

The current VAMDC project presents attractive possibilities for
the OP/IP data activities, particularly with respect to data
exchange and preservation and database upgrading and
integration. One of the first tasks is to transpose the
TIPTOPbase DBMSs, originally developed in fortran, to an
open-source SQL package, namely MySQL. This work is currently
in progress and will facilitate database maintenance and
integration within the virtual federation, apart from making
data manipulation more robust and bug-free and globalizing its
query possibilities.

Data preservation is a relevant issue as most OP/IP members are
retiring or close to retirement. Thus, the atomic data sets
must be kept well-documented with metadata to ensure not only
their future usage but also their identity for the new
generation of search engines. The VAMDC is defining and
establishing standard metadata and XML schemata which need to
be adopted by the OP/IP. Furthermore, with regards to data
activities, we have had very good experience in the past
working with data centers (i.e. CDS) and supercomputer
facilities (i.e. OSC, CeCalCULA), and perhaps these alliances
should be cultivated even further in upcoming developments.

Within the VAMDC time scale, it is timely to complete and
upgrade TIPTOPbase with new data. For instance, the original OP
project neglected some chemical elements such as P, Cl, K and
astrophysically deficient members of the iron group which could
now be calculated and included in TOPbase. Also, the IP should
perhaps reconsider treating the electron impact excitation of
Ni in a similar systematic manner as Fe. With respect to
radiative data and in order to generate accurate massive data
sets, the atomic structure code AUTOSTRUCTURE can now be used
to compute bound--bound relativistic $A$-values for transitions
involving empirically corrected levels with high principal
quantum numbers. Furthermore, with the recent inclusion of the
Breit operators in BPRM, a revision of the opacities in
intermediate coupling could perhaps be envisaged.

There is a final aspect that we must refer to, namely that of
data-intensive application deployment. As atomic data sets
become more voluminous, effective data mining will have to
depend on innovative methods and computer tools. It will become
{\em database-centric} where the application is now run close
to the data residence rather than downloading the data sets to
the place where the application is being run. At present atomic
data-intensive applications, e.g. astrophysical spectra
modeling codes (XSTAR\footnote{\tt
http://heasarc.gsfc.nasa.gov/docs/software/xstar/xstar.html},
CLOUDY\footnote{\tt http://www.nublado.org/} and
TLUSTY\footnote{\tt http://nova.astro.umd.edu/}), are usually
downloaded from a web site, installed locally and run on
sequential processors. This scheme is limited by database
volume and maintenance, and is difficult to adapt and tune to
the new distributed (e.g. the grid) and virtual (e.g. the
cloud) computing environments. Thus, new approaches for
application deployment must be devised such as workflows (e.g.
Taverna\footnote{\tt http://www.taverna.org.uk/}), grid portals
(e.g. GENIUS\footnote{\tt
http://egee.cesnet.cz/en/user/genius.html}) and virtual
machines (e.g. VMware\footnote{\tt http://www.vmware.com/}).
Within the VAMDC project and with NASA support, we are using
XSTAR as a case study of application deployment in the new
cyber-infrastructure.


\section*{Concluding remarks}

In our review of the data activities of the OP and IP, there
are a few points we would like to briefly develop further.
Among them would be the remarkable duration of these
international collaborations which have spanned more that 25
years of active data production. Rather than praising their
stability and cohesion, we would just like to emphasize that
the computing of accurate data basically takes time and
experience. Moreover, one would expect the data produced by
these large-scale efforts to be definitive, but as discussed
here, there is always room for data refinement, for upgrading
the unavoidable approximations, for new emerging discrepancies
and for periodic revisions. These are some of reasons behind
their on-going status. With respect to addenda or revisions, we
have suggested some firsthand cases, but the present solar
abundance problem is indeed an incentive for re-examining the
radiative properties, particularly those involving the inner
shells, of some of the more complex ionic species. In any case,
as shown by the OP, IP and OPAL, relevant scientific problems
can actually be solved by curating and analyzing large volumes
of data, and with the breathtaking evolution of ICT, these
approaches are likely to dominate in the near future.

As we have illustrated here with the OP/IP, large data
productions are usually carried out by global collaborations,
i.e. multinational science groups that generate large volumes
of data which are geographically distributed and maintained
only during project life cycle. It is imperative to plan data
preservation policies and protocols such that these efforts are
not wasted in the long run. The increasing awareness to
preserve and share data for use and reuse in knowledge creation
has to support and promote the emerging Open Data Movement and
its actions. As mentioned previously, we are confident that the
current initiatives in the A\&M community, e.g. the VAMDC, will
seriously look into this problem. An initial step in the right
direction has been the prompt release of an XML schema (XSAMS)
for A\&M data exchange. What is required now is to improve and
normalize metadata identification, and since A\&M data are used
in different fields, key aspects are simplicity, flexibility
and semantic inter-operability.

It is also interesting to point out the existing dichotomy of
supercomputer centers and data centers. In our data activities
in the OP/IP, we have worked successfully with both entities
where their support, technical advice and security consulting
were always invaluable, but they are in themselves very
different in nature and in the services they provide. We are of
the opinion that the essentials of the new cyber-infrastructure
will promote the emergence of virtual distributed environments
where high-performance computing will take place close to the
data repositories. Thus, both supercomputer and data centers
will have to evolve and perhaps merge, and the deployment of
data-intensive applications will engage innovative schemes that
could be somewhat disconcerting to both data producers and
users.

In some of the data activities of the OP/IP  we were lucky to
collaborate with a first-class computer scientist who supported
our database developments. Due to the complexities of the new
cyber-infrastructure, which are beyond most computational
physicists, the reliance on skilled computer scientists and
interdisciplinary teams is likely to become vital in data
dissemination. This would require a scale of funding well above
the ones that were tapped during the course of the OP and IP,
but it appears that funding agencies are now more aware of the
importance of data projects.

Finally, we consider that the current trend for data
dissemination and preservation points to the extensive use of
DDCEs in conjunction with data factories, instruments and
sensors. As previously mentioned, a DDCE is a set of services
for capturing, curating, disseminating and preserving data from
research projects conducted by knowledge communities such as
those we are currently developing within the VAMDC project.


\section*{Acknowledgements}

For the writing of the present review, ample use was made of
the SAO/NASA Astrophysics Data System (ADS). I am indebted to
Natalya Critchley and Keith Butler (Universit\"ats-Sternwarte
M\"unchen, Germany, and OP/IP member) for going over the
typescript in detail. In the context of the OP/IP data
activities, the long-term support of the Centre de Donn\'ees
astronomiques de Strasbourg (CDS) and the Ohio Supercomputer
Center (OSC) is gratefully acknowledged, specially the
technical expertise, guidance and patience of Fran\c{c}ois
Ochsenbein (CDS) and Paul Buerger (OSC). The VAMDC is funded by
the European Union within the FP7 initiative ``Research
Infrastructures -- INFRA-2008-1.2.2 -- Scientific Data
Infrastructures". The Venezuelan node has been supported by
Corporaci\'on Parque Tecnol\'ogico de M\'erida and by EELA-2
(E-science grid facility for Europe and Latin America).



\end{document}